# Improving performance of CNN to predict likelihood of COVID-19 using chest X-ray images with preprocessing algorithms

*Morteza Heidari[1], Seyedehnafiseh Mirniaharikandehei[1], Abolfazl Zargari Khuzani[2], Gopichandh Danala[1], Yuchen Qiu[1], Bin Zheng[1]

[1]School of Electrical and Computer Engineering, University of Oklahoma, Norman USA

[2]Department of Electrical and Computer Engineering, University of California Santa Cruz, Santa Cruz, USA

*correspondence: morteza.heidari@ou.edu

**Abstract-** As the rapid spread of coronavirus disease (COVID-19) worldwide, chest X-ray radiography has also been used to detect COVID-19 infected pneumonia and assess its severity or monitor its prognosis in the hospitals due to its low cost, low radiation dose, and wide accessibility. However, how to more accurately and efficiently detect COVID-19 infected pneumonia and distinguish it from other community-acquired pneumonia remains a challenge. In order to address this challenge, we in this study develop and test a new computer-aided diagnosis (CAD) scheme. It includes several image pre-processing algorithms to remove diaphragms, normalize image contrast-to-noise ratio, and generate three input images, then links to a transfer learning based convolutional neural network (a VGG16 based CNN model) to classify chest X-ray images into three classes of COVID-19 infected pneumonia, other community-acquired pneumonia and normal (non-pneumonia) cases. To this purpose, a publicly available dataset of 8,474 chest X-ray images is used, which includes 415 confirmed COVID-19 infected pneumonia, 5,179 community-acquired pneumonia, and 2,880 non-pneumonia cases. The dataset is divided into two subsets with 90% and 10% of images in each subset to train and test the CNN-based CAD scheme. The testing results achieve 94.0% of overall accuracy in classifying three classes and 98.6% accuracy in detecting Covid-19 infected cases. Thus, the study demonstrates the feasibility of developing a CAD scheme of chest X-ray images and providing radiologists useful decision-making supporting tools in detecting and diagnosis of COVID-19 infected pneumonia.

# 1. INTRODUCTION

From the end of 2019, a new coronavirus, namely COVID-19, was confirmed in human bodies as a new category of diseases that cause dangerous respiratory problems, heart infection, and even death. This virus spreads through the air, object surface, and even person to person in an exponential way. In order to more effectively control the situation of virus spread and treat patients to reduce mortality rate, medical images can play an important role [1]. In current clinical practice, 2D chest X-ray radiography and 3D computed tomography (CT) are two recommended imaging modalities to detect COVID-19, assess its severity and monitor its prognosis (or response to the treatment). In these two imaging modalities, although CT can achieve higher sensitivity read by radiologists, chest X-ray radiography has clinical application advantages including low cost, low radiation dose, easy-to-operate and wide accessibility in general or community hospitals [2]. However, reading high volume of chest X-ray images to detect subtle COVID-19 infected pneumonia and/or distinguish it from other community-acquired pneumonia by general radiologists in the community hospitals may be a difficult and time-consuming task due to many common similarities between pneumonia infected by COVID-19 and caused by other infections like (influenzas). Thus, this is a clinical challenge faced by the radiologists in the clinical practice of this pandemic [3].

In order to address this challenge, developing computer-aided detection or diagnosis (CAD) schemes or systems based on medical image processing and machine learning (i.e., deep learning) algorithms has been attracting broad research interest, which aims to automatically analyze characteristics related to different diseases and then provide radiologists a valuable decision-making supporting tool for a more reliable and accurate detection and diagnosis of COVID-19 infected pneumonia. To this aim, we need to apply image preprocessing algorithms to do image quality enhancement and segmentation, identify and compute image textural features with high association to the diverse diseases, train and develop multiple-feature based machine learning models to detect and classify cases. Due to difficulty to identify and segment subtle pneumonia related disease patterns on 2D chest X-ray images, developing CAD schemes based on deep learning algorithms without segmentation of suspicious disease pattern or regions can be more efficient and probably more reliable than the use of the classical machine learning based methods (i.e., decision trees or support vector machine). This is because that deep learning structures can drive more abstract features in their hidden deep layers, which is impracticable to achieve with classical handcraft feature extraction approaches. In addition, some extracted handcraft features not only have large variance that is not effective in improving the accuracy of the proposed technique, but also have an adverse effect on the final classification results. While in the end-to-end convolution neural network (CNN) based schemes, the algorithm optimizes the hidden features in different layers to the targeted purpose under numerous training iterations automatically.

In the deep learning research field, many models have been developed and applied in different applications, including CAD schemes of medical images for segmentation of regions of interest (ROIs) and detection or classification of various diseases [4, 5]. Recently, several research groups reported their preliminary CAD studies in COVID-19 conditions. For example, one study [6] introduced a COVD-19 classification system based on feature extraction using a deep learning algorithm to feed them into a support vector machine (SVM) to perform disease classification. Several CNN models have investigated in this study and the combination of the Resnet50 for feature extraction plus SVM classifier design yields statistically better performance than the other models. Using a test dataset of 50 2D chest X-ray images, the best accuracy reported is 95.38% to classify between COVID-19 and non-COVID-19 cases. In other studies, Fei et al. [7] performed a deep learning based system on chest CT scan images to automatically segment and quantify the infected area by COVID-19. VB-Net is utilized for CNN, and a dataset of 249 patients is used for training and 300 cases for validation. A Dice similarity of 91.6% between the automatic and manual segmentation is reported for their validation performance. Ioannis et al. in [8] presented an algorithm for automatic COVID-19 detection from chest X-ray images, which is mainly based on transfer learning ideas. To this purpose, the most popular CNN systems are adopted to analyze. The result shows that VGG and MobileNet based CNN systems yield classification accuracy of 96.78%, which is performed superiorly in comparison with the others. Butt et al. in [9] chose a local-attention mechanism model for image feature extraction, and ResNet as a classical CNN system to classify chest CT images depicting COVID-19 with an overall accuracy of 86.7%. Wang et al. in [10] also used a transfer learning based CNN deep learning model to classify COVID-19 using CT images. Pre-trained inception is selected as CNN to establish the algorithm, and a random technique for the region of interest selection (ROI) is proposed for feature extraction. On a test set of 453 CT images, the study reports 82.9% classification accuracy.

We in this study utilize a VGG16 deep neural network and implement it to a new CAD scheme of chest X-ray radiography images to detect and classify images into 3 classes, namely COVID-19 infected pneumonia, the community-acquired other viral pneumonia, and normal (non-pneumonia) cases. Comparing to many other deep learning models, we found that the VGG16 model has the following advantages for this application. First, in the previous studies, VGGNet has demonstrated its superior and stable performance in many engineering applications using diverse datasets. Second, in the competition using the ImageNet dataset involving 14 million images, it scored the first place on image localization task and second place on image classification task in 2014. It means that the parameters of this CNN model are very well trained using ImageNet, which makes it suitable for transfer learning. In this study, to avoid and control the risk of overfitting the deep learning CNN using the limited chest X-ray image dataset, we apply a transfer learning method. A pre-trained VGG16 based on the ImageNet dataset is selected since it enables to extract general shape, structure, and edge related image features in the primarily front hidden layers. The chest X-ray images are applied to do fine-tuning of the limited parameters at the last block of 4 layers of the network. In order to optimally apply the pre-trained VGG16 model in this application, we employ several image pre-processing steps or algorithms on the chest X-ray images. These steps first detect and remove diaphragm region on the image. Next, low pass filter and histogram equalization are applied to the images to reduce image noise and normalize the contrast-to-noise ratio of the images. Then, three different versions of the image are generated and used to feed into three channels of VGG16 model that is originally trained by color images with three input channels (RGB). In this way, we can fully use the capability of the VGG16 model that is pre-trained using color images of the Image-Net database. Finally, we fine-tune the CNN using the pre-processed X-ray images. We will also investigate and compare the utility and advantages of performing initial segmentation and filtering process to achieve better classification performance in detecting cases with COVID-19 infection.

## 2. MATERIALS AND METHOD

### 2.1 Dataset

In this study, we utilize a dataset of chest X-ray radiography (CXR) images assembled from the publicly available medical repositories [11-13]. The dataset was created and examined by the Allen Institute for AI in partnership with the Chan Zuckerberg Initiative, Microsoft Research, Georgetown University's Center for Security and Emerging Technology, and the National Library of Medicine - National Institutes of Health, in coordination with The White House Office of Science and Technology Policy. The dataset includes 8,474 2D X-ray images in the posteroanterior (P.A.) chest view. Among them, 415 images depict with the confirmed COVID-19 disease, 2,880 normal (non-pneumonia) cases, and 5,179 with other community-acquired bacterial and viral pneumonia.

### 2.2 Image Pre-Processing

Figure 1 (a to c) shows examples of three chest X-ray images related to each of the three classes in the dataset. As it is evident in the posteroanterior view, the bottom part of the images includes diaphragm region with high-intensity or bright pixels that may have a negative effect on distinguishing potential patterns of disease in lung areas leading to an automatic diagnosis. Hence, we add an image pre-processing algorithm to identify and remove the diaphragm region depicting on the image. Specifically, the algorithm detects the maximum (the brightest - $V_{max}$) and minimum (the darkest - $V_{min}$) pixel value of the image, then uses a threshold that equals $T = V_{min} + 0.9 \times (V_{max} - V_{min})$, to segment the original image into a binary image. Next, morphological filters (i.e., open, close, and dilate) are applied to detect different connected components with higher intensity (brightness) in the image. From all detected connected regions, the largest one in the image is selected and removed from the image because the high-intensity value of the pixels in the image is defined as the diaphragm region. This process is applied to all cases in the dataset, and we name it as a "blob discovery" step. One example image of removing the diaphragm region is shown in Figure 1(d).

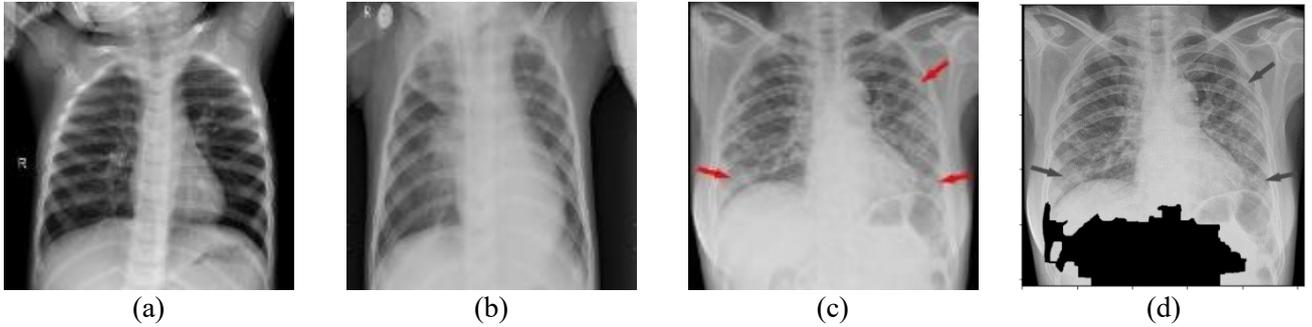

(a)　　　　　　　　　(b)　　　　　　　　　(c)　　　　　　　　　(d)

Figure 1. a) a normal case. b) a case with viral pneumonia c) a case with COVID-19 that the suspicious areas are marked with red arrow. d) The high-intensity object removal process is applied to (c).

In the next step, we convert the segmented grayscale images to 3-channel images suitable for fine-tuning a pre-trained VGG16-based CNN model, which was pre-trained using the ImageNet dataset. To do so, instead of simply copying the original chest X-ray image after removing diaphragm to 3 different channels, we apply two image noise filtering and histogram equalization method to original image. First, since the X-ray images often include additive noise, we apply a bilateral filter that is a non-linear filter and is highly effective at noise removal while preserving textural information compared to the other low pass filters. In other words, this filter analyzes intensity values locally and considers the intensity variation of the local area to replace the intensity value of each pixel with the averaged intensity value of the pixels in the local area. To calculate the weights, we apply a Gaussian low-pass filter in the space domain. This step generates a noise-reduction image. Second, chest X-ray images may have different image contrast or brightness due to the difference in patient body size and/or variation of X-ray dose. In order to compensate such a potentially negative impact, we apply a histogram equation method to normalize image. Thus, applying histogram equalization helps contrast modification that can enhance lung tissue patterns and characteristics associated with COVID-19 infection.

Then, three images namely, (1) the original chest X-ray image after removing diaphragm region ($I_p$), (2) the bilateral filtered image using the Gaussian low-pass filter ($I_b$), and (3) the image normalized using histogram equalization ($I_{eq}$), are fed into three input RGB channels of the VGG16 based CNN model, respectively. Images in all 3 channels are resized to 224×224 pixels in order to fit the VGG16 pre-trained model. Figure 2 shows image pre-processing steps to generate 3 images fed into three channels of the VGG16 based CNN model.

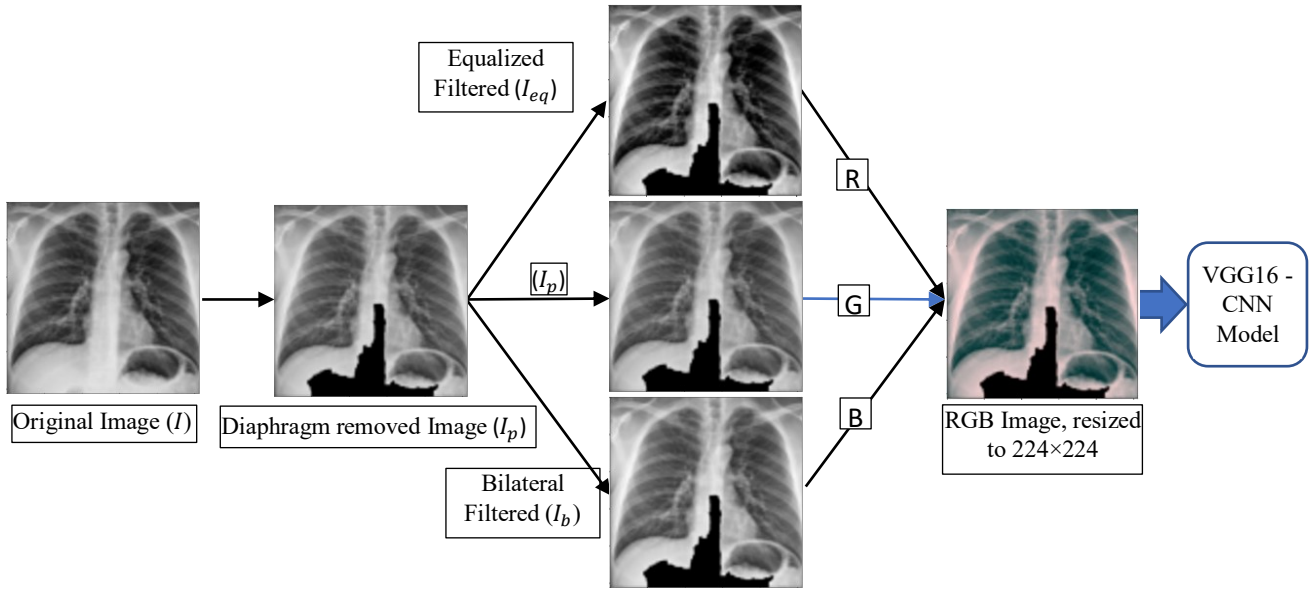

Figure 2. A flow diagram to illustrate image pre-processing steps to generate input of VGG16 based CNN model, where ($I$) is the original Image in the dataset. ($I_p$) is the diaphragm removed image. ($I_{eq}$) is an image after applying histogram equalization on ($I_p$), and ($I_b$) is an image after bilateral filtering on ($I_p$). Each ($I_p$), ($I_b$), and ($I_{eq}$) is fed into one channel to simulate the RGB image used in the model.

## 2.3 Transfer Learning

Usually, when using a dataset of small size, training CNN from scratch is not advisable since CNN has a huge number of parameters that need to be trained and determined, so instead of learning, it may memorize the dataset. To avoid either overfitting or underfitting consequences, it is possible to take advantage of a CNN initially trained on a specific task like classification with a large-scale dataset to overcome the insufficient data [14]. In the initial training of CNN on the large-scale dataset, it has learned to distinguish the most common and outstanding image features like shape, edge, etc. Then, it is suitable for the next phase, which is transfer learning. In this phase, we apply particular modifications like parameter tuning to the pre-trained model to achieve optimal results for the purpose.

The CNN employed for the classification task in this study is a VGG16 model. It is pre-trained in the ImageNet Large Scale Visual Recognition Challenge (ILSVRC), which is a dataset with 14 million images [15]. As shown in Figure 3, the VGG16 model has 16 convolution and fully connected layers in 6 blocks, which include over 138 million trainable parameters. In our transfer learning, the weights between the connected nodes in front or low layers of the network maintain (blocks 1 to 5). The top layers of the VGG16 model (from the beginning of block 6) are cut out and removed. Next, to prepare the CNN for this application purpose, we add one flatten layer, two fully connected layers with 256 and 128 nodes, respectively. In these layers, Rectified Linear Unit (ReLU) [16] is used as their activation function. Then, we add the last classification layer, which uses Softmax as the activation function to build the complete deep learning model to fulfill a three-class classification task. In this way, we have 21,170,755 trainable parameters or connection weights because of the four layers added to the remaining output layer of the base VGG16 model. The complete CNN model is compiled with Adam [17] optimizer with a batch size of 4, max epoch = 200, initial learning rate = $10^{-5}$, and monitoring validation loss for reducing the learning rate every five epochs with a factor of 0.8.

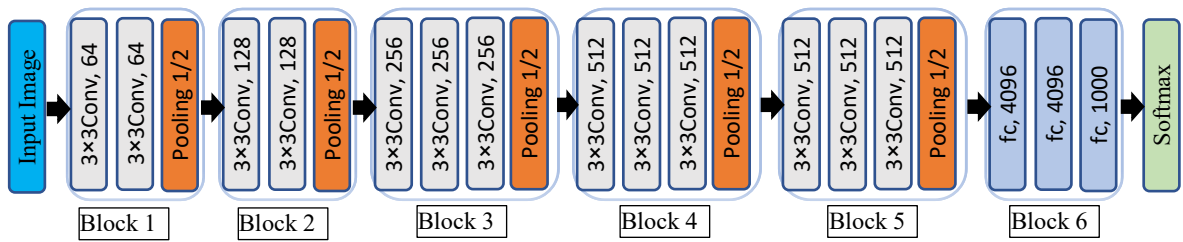

Figure 3. Illustration of the architecture of VGG16 based CNN model.

## 2.4 Model Training and Testing

For training and evaluating the proposed VGG16 based transfer leaning CNN model, we randomly split the entire image dataset into two independent training and testing subsets using a ratio of 90% to 10% cases with respect to the number of cases in each of three classes. Thus, the testing subset includes 42 Covid-19 infected pneumonia, 518 community-acquired pneumonia, and 288 normal (non-pneumonia) cases. The CNN is trained based on the 90% of the cases of the whole dataset involving 415 COVID-19, 2,880 normal, and 5,179 community-acquired pneumonia cases. Besides, in order to reduce the risk of overfitting and determine the optimal training iteration epochs, we also extract 10% of training cases in each class to form an independent validation subset. Then, on the remained training data, we applied shearing, zoom, rotation, width and height shift, and horizontal flip as augmentation techniques commonly used in training deep learning models aiming to increase the size of training samples [18]. Figure 4 shows a complete schematic diagram that illustrates the complete architecture of this VGG16 transfer learning based CNN model, as well as the training, validation, and testing phase.

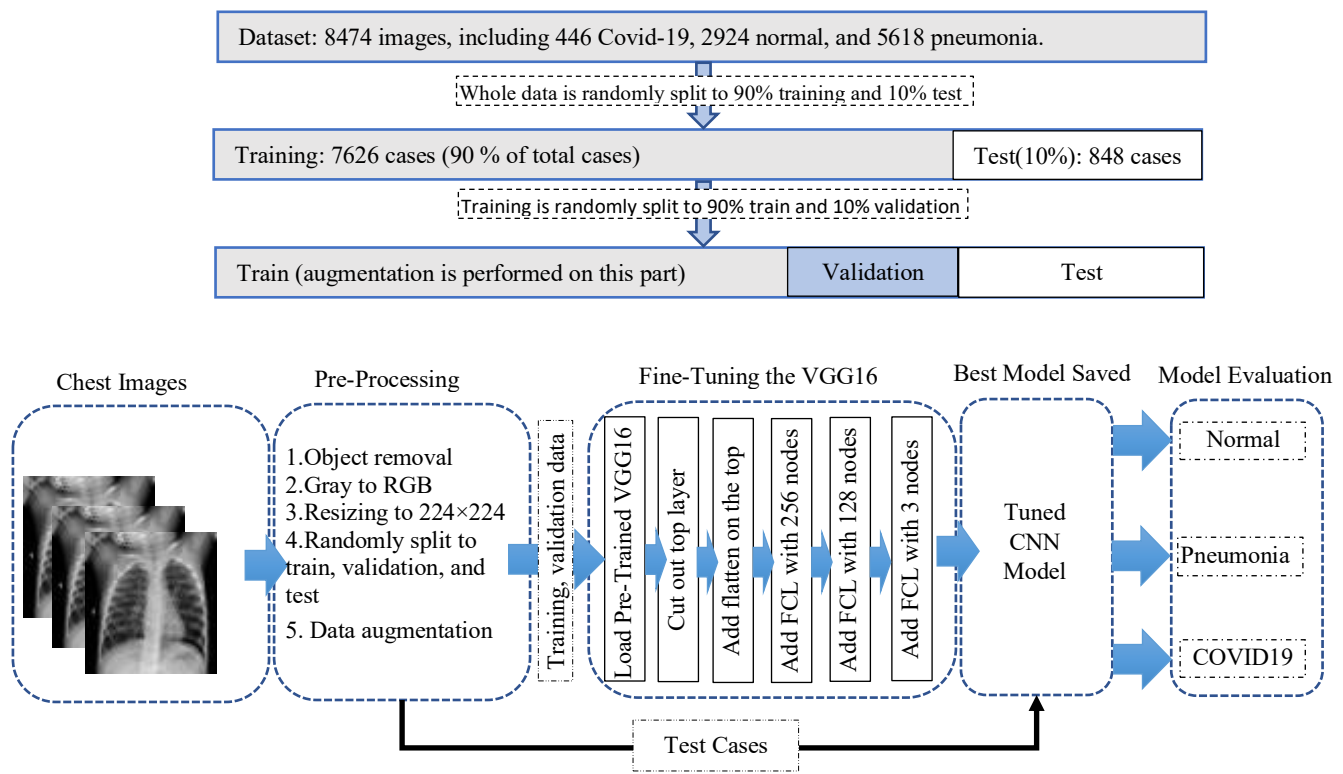

Figure 4. schematic representing training and validation phase of the proposed scheme.

## 2.5 Performance Assessment

While the primary purpose of this study is to detect cases with COVID-19 infection, we analyze and report two different accuracies. The first one is accuracy for a three-class classification problem to distinguish between normal, community-acquired pneumonia, and COVID-19 infected pneumonia cases. Specifically, we compute three accuracy values in detecting images in three classes. We also calculate (1) a macro averaging, which is the average of three accuracy values of three classes without considering the proportion of the number of cases in each class ($A_{mac} = (A_1 + A_2 + A_3)/3$), and (2) a weighted averaging, which is the weighted average of three accuracy values weighted with respect to the proportion of the classes ($A_w = w_1 A_1 + w_2 A_2 + w_3 A_3$), where $A_1, A_2, A_3$ are accuracy values of three classes, while $w_1, w_2, w_3$ are weighting factors of three classes representing the ratios of cases in three classes. Then, for the three-class classification, a confusion matrix is generated from which several evaluation indices, including precision, recall, F1-score, and Cohen's Kappa [19] values are computed to evaluate the performance of the CAD scheme implemented with a trained GVV16 based CNN model. The value of Cohen's kappa coefficient is between zero and one. It is a statistical way to evaluate the robustness of the results. This value shows the possibility of the predicted results occurring by chance. The lower Kappa value shows the more randomness of the results, while the higher value shows a better similarity and higher robustness.

The second accuracy evaluation refers to classification between COVID-19 and non-COVID19 cases (including both normal and community-acquired pneumonia cases). In this circumstance, we compute True Positive (TP) for the cases correctly identified as COVID-19, False Negative (FN) for the COVID-19 cases being incorrectly predicted as normal or community-acquired pneumonia cases, True Negative (TN) for the cases correctly identified as non-COVID-19 cases, and False Positive (FP) for the normal and community-acquired pneumonia cases being incorrectly predicted as COVID-19 by the CNN model. Based on these relationships, the accuracy, sensitivity, specificity, recall, and F1-scores of model prediction or classification are calculated using the following equations.

$$\text{Sensitivity} = TP / (TP+FP) \qquad (1)$$

$$\text{Specificity} = TN / (TN+FP) \qquad (2)$$

$$\text{Accuracy} = (TP+FP) / (\text{Total number of test cases}) \qquad (3)$$

$$\text{Recall} = TP / (TP+FN) \qquad (4)$$

$$\text{F1-score} = (2 \times TP) / (2 \times TP+FP+FN) \qquad (5)$$

## 3. RESULTS

Figure 5(a) presents the trend of training, and validation accuracy of the new transfer learning VGG16 based CNN model as the increase of training iteration epochs during the training process. As shown in the figure, the prediction accuracy of the validation subset varies greatly (with big oscillation) initially, and then gradually converges to a higher accuracy level with much small oscillation. Thus, after epoch 75, validation accuracy is following the training accuracy, which means learning is happening during different epochs. The trend graph also shows that the proposed technique does not suffer significant overfitting or underfitting as two possible happenings in training deep CNN models. For the best validation results, the model is saved and then evaluated using the testing cases. The confusion matrix of the evaluation on the testing cases is shown in figure 5(b), which exhibits that all COVID19 cases are classified correctly with 12 false-positives, including 4 normal cases and 8 community-acquired pneumonia.

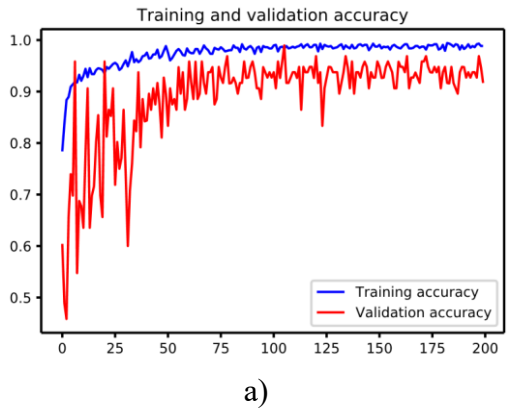 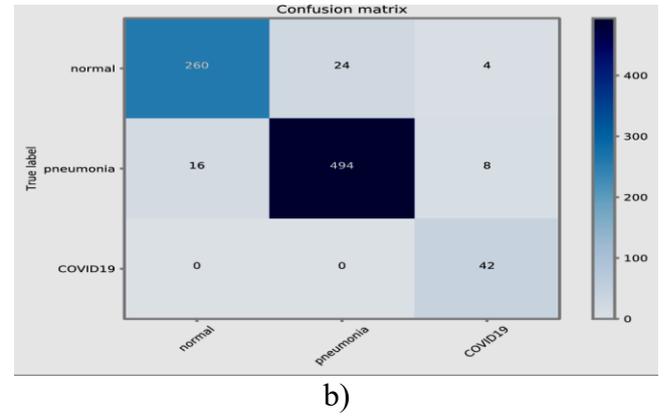

|   |   |
|---|---|
| a) | b) |

Figure 5. (a) Two trend curves show that in the training phase, the train and validation accuracy of the proposed VGG16 based deep CNN model for 200 training iteration epochs. (b) A confusion matrix of the testing cases applying to the best model with the highest validation accuracy in the training phase.

In the next step from the confusion matrix, we compute and determine the precision, recall rate, F1-score, and prediction accuracy of the new VGG16 based CNN model as shown in Table 1. Among 848 testing cases, 796 are correctly detected and classified into three classes. Thus, the overall accuracy is 93.9% (796 / 848) with 95% confidence interval of [0.92,0.96]. In addition, the Cohen's kappa coefficient for the proposed method is 0.88, which confirms the reliability of the proposed approach to train this new deep learning model to do this classification task.

Table 1. Classification report of the proposed method.

|   | Precision | Recall | F1-score | Support cases |
|---|---|---|---|---|
| **Normal** | 0.94 | 0.90 | 0.92 | 288 |
| **Other Pneumonia** | 0.95 | 0.95 | 0.95 | 518 |
| **COVID19** | 0.78 | 1.00 | 0.88 | 42 |
| **Accuracy** | --- | --- | **0.94** | 848 |
| **Macro avg** | 0.89 | 0.95 | 0.92 | 848 |
| **Weighted avg** | 0.94 | 0.94 | 0.94 | 848 |

To future evaluate the performance of the developed CAD scheme with the VGG16 based CNN model in detecting COVID19 infection using chest X-ray images, we place both normal and community-acquired pneumonia images into the negative class and COVID-19 infected pneumonia cases into the positive class. Combining the data in the confusion matrix as shown in Figure 5(b), the CAD scheme yields 100% sensitivity (42/42) and 98.5% specificity (794/806). The overall accuracy is 98.6% (836/848).

Next, to assess the impact of the pre-processing steps namely, removing the diaphragm region in the image ($I_p$) and the application of two pre-processing filters ($I_b$), ($I_{eq}$), we perform simulation results for two other pre-trained VGG16 based CNN models with the same architecture. For the first model, each original X-ray image ($I$) as shown in figure 2 is simply duplicated copied to all three RGB channels of the VGG16 model. In other words, we skip all image pre-processing steps, and the original $I$ is applied to all three channels. Then, the VGG16 based CNN model is trained using the same training and validation process to build an optimal model, and performance is tested using the same testing cases. We name this model "a simple model." For the second model, just phase one of the pre-processing steps, which is removing diaphragm region, as shown in the figure 2, is skipped. Two filters are directly applied to the original image ($I$). Then, the original image and two filtered images are applied to the three input channels of the GVV16 model. The same training, validation and testing steps described above are applied to this model. We name this model "a filter base model". Table 2 shows two confusion matrixes of these two models. Based on these two confusion matrixes, the accuracy of the "simple model" is 88% and Cohen's kappa score of 0.75, while the "filter base model" yields 91% accuracy, and Cohen's kappa score of 0.82.

**Table 2. Confusion matrix of two CNN models with and without filtering on the original X-ray Images. 95% of confidence interval (CI) for the accuracy is shown in the last column.**

|  |  | Normal | Pneumonia | COVID19 | Accuracy | 95% CI |
|---|---|---|---|---|---|---|
| **filter base model** | Normal | 228 | 55 | 5 | 91.0% | [0.89,0.93] |
|  | Pneumonia | 6 | 503 | 9 |  |  |
|  | COVID19 | 1 | 0 | 41 |  |  |
| **simple model** | Normal | 197 | 81 | 10 | 87.6% | [0.85,0.90] |
|  | Pneumonia | 4 | 506 | 8 |  |  |
|  | COVID19 | 1 | 1 | 40 |  |  |

(True Label)

## 4. DISCUSSION

In this study, we developed and tested a novel deep learning CNN model to predict the likelihood of testing chest X-ray images depicting with COVID19 infected pneumonia disease. This study has several unique characteristics as comparing to the previously reported studies in this field and produces several new interesting observations. First, since deep learning CNN model includes a considerable number of parameters that need to be trained and determined, a large and diverse image dataset is required in order to produce robust results [20]. Although we used a relatively large image dataset of 8,474 chest X-ray images, the dataset is unbalanced in three classes of images, and the number of COVID-19 infected pneumonia cases (415) remains small. Thus, in order to build a robust deep learning model, we select a well-trained VGG16 model and apply a transfer learning approach. Specifically, the original VGG16 model includes over 138 million parameters. These parameters have been trained and determined using a large ImageNet database over 14 million images. It is difficult to re-train so many parameters robustly using our dataset of 8,474 images. Thus, we only re-train or fine-tune the limited parameters in the last block of the pre-train VGG16 (as shown in Figure 3) to reduce the risk of overfitting. The study results demonstrate that this transfer learning approach can yield higher performance with the overall accuracy of 93.9% (796/848) in the classification of three classes and 98.6% (836/848) in classifying cases with and without COVID-19 disease, as well as the high robustness with a Cohen's kappa score of 0.88.

Second, unlike the regular color photographs, chest X-ray images are gray-level images. Thus, in order to fully use the pre-trained VGG16-based CNN model, we generate two new gray-level images. Then, instead of applying the original chest X-ray image to the CNN model directly only, three different gray-level images are fed into three different input (RGB color) channels of the CNN model. The two new gray-level images contain additional information to enhance image classification capability. Specifically, in this study, we apply a bilateral filter to generate a noise-reduced image and a histogram equalization method to generate a contrast normalized image. Comparing two approaches of using only original chest X-ray image and three different images as input images to the CNN model, our study results show that using three different input image approach, overall classification accuracy increases 3.2% from 91.0% to 93.9%, and Cohen's kappa score increases 7.3% from 0.82 to 0.88, respectively. The results demonstrate the advantage of using our new approach to fully use three input channels of the CNN model pre-trained using color images.

Third, since in the area of medical imaging, generally disease's pattern are not comparable to the other existing patterns in the image, pre-processing steps are noteworthy [21], so we apply an image pre-processing algorithm to automatically detect and remove the majority part of the diaphragm region from the chest X-ray images. Comparing the approaches with and without removing the diaphragm regions, classification performance of the CNN model changes from 93.9% to 87.6% and 0.88 to 0.74 for the overall classification accuracy and Cohen's kappa coefficients, respectively, which indicate that the overall classification accuracy increases 7.2% and Cohen's kappa coefficient increases 18.9% by removing the majority of diaphragm regions. Thus, although skipping segmentation of the suspicious disease regions of interest is one important characteristic of deep learning, our study demonstrates that applying an image processing and segmentation algorithm to remove irrelevant regions on the image can also play an important role to increase model performance and robustness for the datasets which are not large enough for a CNN based model.

In addition, we observe that applying data augmentation in the training data is also essential. Without data augmentation to increase training dataset size, the overall classification accuracy of the CNN model significantly reduces to around 78%. In summary, we in this paper present a new deep learning model to detect and classify COVID-19 infected pneumonia cases, as well as several unique approaches to optimally train the deep learning model using the limited and unbalanced medical image dataset. The similar learning concept and pre-processing approaches can also be adopted to develop new deep learning models for other medical images to detect and classify other types of diseases (or cancers [22, 23]).

Although the performance of detecting and classifying COVID-19 infected pneumonia cases is encouraging, we recognize the limitation of this study. First, although we used a publicly available dataset of 8,474 cases, including 415 COVID-19 cases, due to the diversity or heterogeneity of COVID-19 cases, the performance, and robustness of this deep learning CNN model based CAD scheme needs to be further tested and validated using other large and diverse image databases. Second, this study only investigates and tests two image pre-processing methods to generate two filtered images, which may not be the best or optimal methods. New methods should also be investigated and compared in future studies. Third, to further improve model performance and robustness, it also needs to develop new image processing and segmentation algorithms to more accurately remove diaphragm and other regions outside lung areas in the images. Therefore, more research work is needed to overcome these limitations in the future studies.

## ACKNOWLEDGEMENT

This work is support in part by the grant from National Cancer Institute (R01 CA197150). The authors also thank the research support from Stephenson Cancer Center, University of Oklahoma, which helps establishment of our Computer-aided Diagnosis Laboratory.